\documentclass[a4paper,11pt]{article}
\usepackage{pos}

\usepackage{hyperref}

\newcommand{\nn}{\nonumber}
\newcommand{\lsim}{\mathrel{\mathop{\kern 0pt \rlap
  {\raise.2ex\hbox{$<$}}}
  \lower.9ex\hbox{\kern-.190em $\sim$}}}
\newcommand{\gsim}{\mathrel{\mathop{\kern 0pt \rlap
  {\raise.2ex\hbox{$>$}}}
  \lower.9ex\hbox{\kern-.190em $\sim$}}}

\newcommand{\be}{\begin{equation}}
\newcommand{\ee}{\end{equation}}
\newcommand{\bea}{\begin{eqnarray}}
\newcommand{\eea}{\end{eqnarray}}

\title{The scalar potential of the 331 model: theoretical constraints}

\author*[a]{Antonio Costantini}

\affiliation[a]{INFN - Sezione di Bologna,\\
  Via Irnerio 46, 40126 Bologna, Italy}

\emailAdd{antonio.costantini@bo.infn.it}

\abstract{We discuss the main features of the scalar sector of a class of BSM models with enlarged gauge symmetry, the so called 331 Models. The theoretical constraints on the scalar potential such as unitarity, perturbativity and boundedness-from-below, are presented, together with the analytical exact digitalization of the scalar sector. The phenomenology of exotic scenarios predicted by the 331 Models can be tested in light of these theoretical constraints.}

\FullConference{%
  40th International Conference on High Energy physics - ICHEP2020\\
  July 28 - August 6, 2020\\
  Prague, Czech Republic (virtual meeting)
}

\begin{document}
\maketitle

\section{Introduction}\label{sec:intro}
\noindent
The $331$ Model~\cite{Singer:1980sw,Valle:1983dk,Pisano:1991ee,Frampton:1992wt,Foot:1994ym,Hoang:1995vq} is an extension of the Standard Model (SM) where the non-abelian gauge group $SU(2)$ of the electroweak symmetry is promoted to an $SU(3)$. This assumption redefines the SM hypercharge as $\mathbb{Y}=\beta_Q \mathbb{T}^8+X\mathbb{I}$. When $\beta_Q$, which is a free parameter of the model, is not specified, the setup is called ``general $331$ Model''. Concerning its particle spectrum, any realisation of the $331$ Model is accompanied by a rich variety of beyond-the-SM (BSM) particles that allows for heavy and potentially exotic states. We present the study of the theoretical constraint on the scalar potential of the $331$ Model. Specifically, we derive the necessary and sufficient conditions for its boundedness-from-below (BFB), we give the analytic expressions of the eigenvalues of the scattering matrix related to perturbative-unitarity and we provide also the analytic expression of the potential's parameter in terms of the physical masses of the scalars.

\section{The scalar sector of the general $331$ Model}\label{sec:model}
\noindent
The general $331$ Model represents a class of SM extensions containing the enlarged gauge group $SU(3)_c\times SU(3)_L\times U(1)_X$. Several specific versions can be obtained by a particular choice of the $\beta_{Q}$ parameter. In the general $331$ Model, the electroweak symmetry breaking is realised by scalars accommodated within three triplets of $SU(3)_L$ 
\begin{equation}\label{eq:scalars}
\chi=\left(
\begin{array}{c}
\chi^A\\
\chi^B\\
\chi^0
\end{array}
\right)
,\quad\rho=\left(
\begin{array}{c}
\rho^+\\
\rho^0\\
\rho^{-B}
\end{array}
\right)
,\quad\eta=\left(
\begin{array}{c}
\eta^0\\
\eta^-\\
\eta^{-A}
\end{array}
\right)
\end{equation}\label{eq:charges}
where each triplet belongs to $(1,3,X)$ with
\be
X_\chi=\beta_{Q}/\sqrt3,\quad X_\rho=1/2-\beta_{Q}/(2\sqrt3),\quad X_\eta=-1/2-\beta_{Q}/(2\sqrt3).
\ee
In addition to neutral and singly charged states, there are fields with charge
\begin{equation}\label{eq:ABcharges}
Q^A=\frac{1}{2}+\frac{\sqrt3}{2}\beta_{Q}\;,\quad Q^B=-\frac{1}{2}+\frac{\sqrt3}{2}\beta_{Q}.
\end{equation}
The scalar potential reads 
\begin{align}\label{pot}
V&= m^2_1\, \rho^*\rho+m^2_2\,\eta^*\eta+m^2_3\,\chi^*\chi +\sqrt2 f_{\rho\eta\chi} \rho\, \eta\, \chi\nn\\
&+\lambda_1 (\rho^*\rho)^2+\lambda_2(\eta^*\eta)^2+\lambda_3(\chi^*\chi)^2\nn\\
&+\lambda_{12}\rho^*\rho\,\eta^*\eta+\lambda_{13}\rho^*\rho\,\chi^*\chi+\lambda_{23}\eta^*\eta\,\chi^*\chi\nn\\
&+\zeta_{12}\rho^*\eta\,\eta^*\rho+\zeta_{13}\rho^*\chi\,\chi^*\rho+\zeta_{23}\eta^*\chi\,\chi^*\eta,
\end{align}
To derive the theoretical constraint on the scalar potential in general one has to analyse the behaviour of the highest powers of the fields, \emph{i.e.} the properties of the quartic couplings of the ultraviolet-complete theory.
For this purpose, it is convenient to parameterise a triplet in the following form:
\begin{equation}
\Phi_i = \sqrt{r_i} e^{i\,\gamma_i}\left(
\begin{array}{c}
\sin a_i \cos b_i\\
e^{i\,\beta_i} \sin a_i \sin b_i\\
e^{i\,\alpha_i}\cos a_i
\end{array}
\right),\qquad i=1,2,3,
\end{equation}
where the fields are complex numbers, $a_i,\,b_i,\,\beta_i,\,\gamma_i$ and $\alpha_i$ are angular parameters, and $r_i$ is the radial part of the field. For the sake of convenience, the following quantity is introduced:
\begin{align}\label{zijgen}
\tau_{ij} & = \left(\Phi_i^\dagger \Phi^{\phantom{\dagger}}_i\right)\, \left(\Phi_j^\dagger \Phi^{\phantom{\dagger}}_j\right) - \left(\Phi_i ^\dagger \Phi^{\phantom{\dagger}}_j\right)\, \left(\Phi_j^\dagger \Phi^{\phantom{\dagger}}_i\right),
\end{align}
where the non-negativity of $\tau_{ij}$ is ensured by the Cauchy--Schwarz inequality. The quartic part of the scalar potential can be written 
in terms of 
a \textit{radial} and an \textit{angular} block:
\begin{align}
V^{(4)} &= V_R + \zeta^\prime_{12}\tau^{\phantom{\prime}}_{12}+\zeta^\prime_{13}\tau^{\phantom{\prime}}_{13}+\zeta^\prime_{23}\tau^{\phantom{\prime}}_{23} = V_R + V_A,
\end{align} 
where the $\zeta$ parameter were conveniently traded with $\zeta^\prime_{ij} = - \zeta^{\phantom{\prime}}_{ij}$ and the radial part reads
\begin{align}
V_R&=\lambda_1 (\rho^*\rho)^2+\lambda_2(\eta^*\eta)^2+\lambda_3(\chi^*\chi)^2\nn\\
&+\lambda^\prime_{12}\rho^*\rho\,\eta^*\eta+\lambda^\prime_{13}\rho^*\rho\,\chi^*\chi+\lambda^\prime_{23}\eta^*\eta\,\chi^*\chi,
\end{align}
with $\lambda^\prime_{ij} = \lambda^{\phantom{\prime}}_{ij}+\zeta^{\phantom{\prime}}_{ij}$.

\section{Boundedness from below}\label{sec:bfb}
\noindent
We present the results of the BFB analysis detailed in \cite{Costantini:2020xrn}. The presence of the angular part of the scalar potential is an essential point of the analysis. For this reason the scenario with $V_A\neq0$ requires a dedicated focus. The BFB of the radial part of the scalar potential is obtained by imposing the co-positivity constraints \cite{Hadeler1983,Kannike:2012pe,Faro:2019vcd} on the matrix $Q_{ij}$, defined by
\begin{equation}
V_R \equiv Q_{ij}r_i r_j.
\end{equation}
This is the set of necessary and sufficient conditions for the BFB of the potential for the case $\zeta^\prime_{12}=\zeta^\prime_{13}=\zeta^\prime_{23}=0$. A good strategy to get rid of the angular information of $V_{A}$ is to search for an ``angularly minimised'' scalar potential with radial dependence only. In applying this procedure one has to consider separately the case where at least one of the $\zeta^\prime$ is zero and where all the $\zeta^\prime$ are different from zero. In the former case the BFB conditions call for co-positivity constraints applied on the new matrices $\widetilde{Q}_k$ defined by
\begin{equation}
V_{R} + \mbox{min}(V_{A})^{T}_{k} = \widetilde{Q}_{k}^{ij} r^{\phantom{\prime}}_i r^{\phantom{\prime}}_j,\,\quad k=1,\ldots,4.
\end{equation}
with the ``trivial'' minima of $V_A$ given by
\begin{align}
 \mbox{min}(V_A)^{T}_{1}&=\zeta^\prime_{12}\,r^{\phantom{\prime}}_1 r^{\phantom{\prime}}_2 + \zeta^\prime_{23}\,r^{\phantom{\prime}}_2 r^{\phantom{\prime}}_3, \\
 \mbox{min}(V_A)^{T}_{2}&=\zeta^\prime_{13}\,r^{\phantom{\prime}}_1 r^{\phantom{\prime}}_3 + \zeta^\prime_{23}\,r^{\phantom{\prime}}_2 r^{\phantom{\prime}}_3, \\ 
 \mbox{min}(V_A)^{T}_{3}&=\zeta^\prime_{12}\, r^{\phantom{\prime}}_1 r^{\phantom{\prime}}_2 + \zeta^\prime_{13}\, r^{\phantom{\prime}}_1 r^{\phantom{\prime}}_3,\\
 \mbox{min}(V_A)^{T}_{4}&=\zeta^\prime_{12}\, r^{\phantom{\prime}}_1 r^{\phantom{\prime}}_2 + \zeta^\prime_{13}\, r^{\phantom{\prime}}_1 r^{\phantom{\prime}}_3 + \zeta^\prime_{23}\, r^{\phantom{\prime}}_2 r^{\phantom{\prime}}_3.
\end{align}
In the latter case the matrix $\widehat{Q}_{ij}$ defined by
\begin{equation}
V_{R} + \mbox{min}(V_{A})^{NT} = \widehat{Q}^{ij} r_i r_j,
\end{equation}
is also required to fulfil the co-positivity criterion, once the transformation described in~\cite{Faro:2019vcd} is applied. In this case the ``non-trivial'' minimum of the angular part is
\begin{equation}
\mbox{min}(V_{A})^{NT} = \frac{\zeta^\prime_{12}\zeta^\prime_{13}\zeta^\prime_{23}}{4}\left(\frac{r_1}{\zeta^\prime_{23}}+\frac{r_2}{\zeta^\prime_{13}}+\frac{r_3}{\zeta^\prime_{12}}\right)^2.
\end{equation}

\section{Perturbative Unitarity}\label{sec:unitarity}
\noindent
The methodology to obtain perturbative unitarity constraints on the SM was described for the first time in~\cite{Lee:1977eg}:
\ all the possible $2 \rightarrow 2$ processes with a given total charge $\mathcal{Q}$ should be considered and the corresponding amplitudes arranged in a scattering matrix. The perturbative unitarity condition imposes then that the real part of the largest eigenvalue of this matrix should not exceed $1/2$. In the general $331$ Model, the scalars of Eq.~\ref{eq:scalars} can have charges $0$, $\pm1$, $\pm Q^A$ and $\pm Q^B$, where $Q^A$ and $Q^B$ take different values depending on the specific realisation of the $331$ Model, \textit{i.e.} of the value of $\beta_{Q}$ (see Eq.~\ref{eq:ABcharges}). It follows that there are $13$ scattering matrices, corresponding to the initial total charge of the $2 \rightarrow 2$ processes 
\begin{align}\label{eq:charge_scattering_matrices}
\mathcal{Q}&=0,\,1,\,2,\,Q^A,\,Q^B,\,Q^A+1,\,Q^B+1,\,Q^A-1,\,Q^B-1,\, \nonumber \\
&\quad Q^A+Q^B,\,Q^A-Q^B,\,2Q^A,\,2Q^B.
\end{align} 
The final condition for the perturbative unitarity is then
\begin{equation}
 |\mathbf{a}|\leq\frac{1}{2}
\end{equation}
where $\mathbf{a}$ identifies all eigenvalues. Their form is shown in the following list
\begin{align}
\mathbf{a}=\Big\{&\frac{\lambda_i}{8\pi},\frac{\lambda_{ij}}{16\pi},\frac{\lambda_{ij}\pm\zeta_{ij}}{16\pi},\frac{\lambda_{ij}+2\zeta_{ij}}{16\pi},\frac{\lambda_i+\lambda_j\pm\sqrt{(\lambda_i - \lambda_j)^2 + \zeta_{ij}^2}}{16\pi},\nn\\
 &\frac{\mathcal{P}_1^3 (\lambda_m,\lambda_{mn},\zeta_{mn})}{32\pi},\frac{\mathcal{P}_2^3 (\lambda_m,\lambda_{mn},\zeta_{mn})}{32\pi}\Big\}
\end{align}
where $\mathcal{P}_{1,2}^3$ are the solutions of third-grade polynomials given by
\begin{align}
& \sum_{i,j,k=1}^3\left[ \frac{x^3}{27}-\frac{4}{9} \lambda_ix^2
+\Big(2 \left( 4 \lambda_i \lambda_j-\zeta_{ij}^2\right)x-\frac{8}{3}\big(\zeta_{ij} \zeta_{ik} \zeta_{jk}-3 \lambda_i \zeta_{jk}^2\right.\nn\\
&\left.\qquad\qquad\qquad\qquad\qquad\qquad\qquad\quad+4 \lambda_i \lambda_j \lambda_k\big)\Big)\left(\varepsilon_{ijk}\right)^2\right],\\
 &\sum_{i,j,k=1}^3\left[ \frac{x^3}{27}
-\frac{16}{9} \lambda_ix^2
+\Big(2 ( 64 \lambda_i \lambda_j-(3 \lambda_{ij}+\zeta_{ij})^2)x\nn \right.\\
&\left.\qquad\qquad-\frac{8}{3}\big(
\zeta_{ik} \zeta_{jk} (9 \lambda_{ij}+\zeta_{ij})+27 \lambda_{ij} \lambda_{ik} (\lambda_{jk}+\zeta_{jk})\right.\nn\\
&\left.\qquad\qquad+4 \lambda_{i} \big(64 \lambda_{j} \lambda_{k}-3 (3 \lambda_{jk}+\zeta_{jk})^2\big)
\big)\Big)\left(\varepsilon_{ijk}\right)^2\right],
\end{align}
with $\lambda_{ji}=\lambda_{ij}$, $\zeta_{ji}=\zeta_{ij} $.

\section{Perturbativity}\label{sec:perturbativity}
\noindent
Requesting perturbative unitarity to be respected is necessary but not sufficient to guarantee the correct perturbative behaviour of the model. Perturbativity of the couplings should also be enforced,
\ setting further theoretical constraints on the parameters of the model. These constraints turn out to be especially effective when the couplings of the scalar potential are recast in terms of physical parameters, according to the diagonalisation procedure described in \cite{Costantini:2020xrn}. We can write schematically
\begin{align}
\rm \lambda&=\rm  F_\lambda(\rm m_{h_i},\rm m_a,\rm m_{h^{\pm Q}},\rm  v_j,\rm \alpha_k)\label{eq:lamexp}\\
\rm f_{\rho\eta\chi}&=\rm F_{\rm f_{\rho\eta\chi}}(\rm m_{h_i},\rm m_a\rm ,m_{h^{\pm Q}},\rm v_j,\rm \alpha_k)\\
\rm \zeta&=\rm F_\zeta(\rm m_{h_i},\rm m_a\rm ,m_{h^{\pm Q}},\rm v_j,\rm \alpha_k)\label{eq:zetaexp}
\end{align}
Remarkably, for a mass spectrum that lives above the electroweak VEV, Eqs.~\ref{eq:zetaexp} calls for a certain degree of degeneracy between $m^2_{h_1^\pm}$ and $m^2_{a_1^{\phantom{\pm}}}$. Beyond that, the set of Eqs.~\ref{eq:lamexp} does not provide any general take-home messages. Even if specific benchmark choices could lead to more manageable formulae, a numerical approach is always required to investigate generic scenarios.

\section{Conclusions}\label{sec:conclusions}
\noindent
We present the analysis of the theoretical constraint on the scalar potential of the $331$ Model. We derive the necessary and sufficient conditions for boundedness-from-below, the analytic constraint of perturbativity as well as perturbative unitarity. The Lagrangian parameters were expressed in terms of the physical parameters, namely masses and mixing angles, by means of a systematic diagonalisation of all the mass matrices of the scalar sector. Perturbativity and perturbative unitarity were then discussed in this spirit and maintaining a consistent general approach.


\begin{thebibliography}{99}

\bibitem{Singer:1980sw}
M.~Singer, J.~W.~F. Valle, J.~Schechter, {Canonical Neutral Current Predictions
  From the Weak Electromagnetic Gauge Group $SU(3) \times U(1)$}, Phys. Rev. D22
  (1980) 738.
\newblock \href {https://doi.org/10.1103/PhysRevD.22.738}
  {\path{doi:10.1103/PhysRevD.22.738}}.

\bibitem{Valle:1983dk}
J.~W.~F. Valle, M.~Singer, {Lepton Number Violation With Quasi Dirac
  Neutrinos}, Phys. Rev. D28 (1983) 540.
\newblock \href {https://doi.org/10.1103/PhysRevD.28.540}
  {\path{doi:10.1103/PhysRevD.28.540}}.

\bibitem{Pisano:1991ee}
F.~Pisano, V.~Pleitez, {An $SU(3) \times U(1)$ model for electroweak interactions},
  Phys. Rev. D46 (1992) 410--417.
\newblock \href {http://arxiv.org/abs/hep-ph/9206242}
  {\path{arXiv:hep-ph/9206242}}, \href
  {https://doi.org/10.1103/PhysRevD.46.410}
  {\path{doi:10.1103/PhysRevD.46.410}}.

\bibitem{Frampton:1992wt}
P.~H. Frampton, {Chiral dilepton model and the flavor question}, Phys. Rev.
  Lett. 69 (1992) 2889--2891.
\newblock \href {https://doi.org/10.1103/PhysRevLett.69.2889}
  {\path{doi:10.1103/PhysRevLett.69.2889}}.

\bibitem{Foot:1994ym}
R.~Foot, H.~N. Long, T.~A. Tran, {$SU(3)_L \otimes U(1)_N$ and $SU(4)_L \otimes
  U(1)_N$ gauge models with right-handed neutrinos}, Phys. Rev. D50~(1) (1994)
  R34--R38.
\newblock \href {http://arxiv.org/abs/hep-ph/9402243}
  {\path{arXiv:hep-ph/9402243}}, \href
  {https://doi.org/10.1103/PhysRevD.50.R34}
  {\path{doi:10.1103/PhysRevD.50.R34}}.

\bibitem{Hoang:1995vq}
H.~N. Long, {The $331$ model with right handed neutrinos}, Phys. Rev. D53 (1996)
  437--445.
\newblock \href {http://arxiv.org/abs/hep-ph/9504274}
  {\path{arXiv:hep-ph/9504274}}, \href
  {https://doi.org/10.1103/PhysRevD.53.437}
  {\path{doi:10.1103/PhysRevD.53.437}}.




\bibitem{Costantini:2020xrn}
A.~Costantini, M.~Ghezzi and G.~M.~Pruna,
{Theoretical constraints on the Higgs potential of the general $331$ model}
Phys. Lett. B 808 (2020), 135638.
\newblock \href{https://arxiv.org/abs/2001.08550} {\path{arXiv:2001.08550}}, \href {https://doi.org/10.1016/j.physletb.2020.135638} {\path{doi:10.1016/j.physletb.2020.135638}}.



\bibitem{Hadeler1983}
K.~P. Hadeler,
  {On copositive matrices}, Linear Algebra and its Applications 49 (1983) 79 -- 89.
\newblock \href {https://doi.org/https://doi.org/10.1016/0024-3795(83)90095-2}
  {\path{doi:https://doi.org/10.1016/0024-3795(83)90095-2}}.
  

\bibitem{Kannike:2012pe}
K.~Kannike, {Vacuum Stability Conditions From Copositivity Criteria}, Eur.
  Phys. J. C72 (2012) 2093.
\newblock \href {http://arxiv.org/abs/1205.3781} {\path{arXiv:1205.3781}},
  \href {https://doi.org/10.1140/epjc/s10052-012-2093-z}
  {\path{doi:10.1140/epjc/s10052-012-2093-z}}.
  

\bibitem{Faro:2019vcd}
F.~S. Faro, I.~P. Ivanov, {Boundedness from below in the $U(1)\times U(1)$
  three-Higgs-doublet model}, Phys. Rev. D100~(3) (2019) 035038.
\newblock \href {http://arxiv.org/abs/1907.01963} {\path{arXiv:1907.01963}},
  \href {https://doi.org/10.1103/PhysRevD.100.035038}
  {\path{doi:10.1103/PhysRevD.100.035038}}.    
  
  

\bibitem{Lee:1977eg}
B.~W. Lee, C.~Quigg, H.~B. Thacker, {Weak Interactions at Very High-Energies:
  The Role of the Higgs Boson Mass}, Phys. Rev. D16 (1977) 1519.
\newblock \href {https://doi.org/10.1103/PhysRevD.16.1519}
  {\path{doi:10.1103/PhysRevD.16.1519}}.  

\end{thebibliography}
\end{document}